\documentclass[aps,preprint,showpacs,floatfix]{revtex4-2}
\usepackage{bm}
\usepackage{amsmath}
\usepackage{epsfig}
\usepackage{rotating}
\usepackage{graphicx}
\usepackage{amssymb}

\begin{document}

\title{Non-linear correlation analysis in financial markets using hierarchical clustering}

\author{J.  E.  Salgado-Hern\'andez}
\affiliation{Instituto de Ciencias F\'isicas,  Universidad Nacional Aut\'onoma de M\'exico,  62210 Cuernavaca,  M\'exico}

\author{Manan Vyas}
\affiliation{Instituto de Ciencias F\'isicas,  Universidad Nacional Aut\'onoma de M\'exico,  62210 Cuernavaca,  M\'exico}
\email{manan@icf.unam.mx}

\begin{abstract}

Distance correlation coefficient (DCC) can be used to identify new associations and correlations between multiple variables. The distance correlation coefficient applies to variables of any dimension,  can be used to determine smaller sets of variables that provide equivalent information,  is zero only when variables are independent,  and is capable of detecting nonlinear associations that are undetectable by the classical Pearson correlation coefficient (PCC).  Hence,  DCC provides more information than the PCC.  We analyze numerous pairs of stocks in S\&P500 database with the distance correlation coefficient and provide an overview of stochastic evolution of financial market states based on these correlation measures obtained using agglomerative clustering.

\end{abstract}

\maketitle

\section{Introduction}

Correlation coefficient is a number which is used to describe dependence between random observations.  Most popular correlation coefficient is the Pearson one which is defined on the interval $[-1, 1]$ \cite{Ma-99}.  For random variables $X$ and $Y$,  with finite and positive variances,  Pearson correlation coefficient (PCC) is defined as $PCC(X,Y) = cov(X,Y)/\sqrt{var(X) \; var(Y)}$.  If Pearson correlation coefficient between two random variables is zero,  it does not necessarily mean that the variables are independent.  Distance correlation coefficient does not suffer from this drawback.  

The distance correlation coefficient (DCC) is a product-moment correlation and a generalized form of bivariate measures of dependency \cite{Sz-09}.  It is a very useful and unexplored area for statistical inference.  The range of the distance correlation is $0 \leq DCC \leq 1$ \cite{Ede-21}.  For two real random variables $X$ and $Y$ with finite variances,  distance correlation coefficient is defined as $DCC(X,Y) = dcov(X,Y)/\sqrt{dcov(X,X) \, dcov(Y,Y)}$.  Here,  the distance covariance $dcov$ is defined in the following way.  Let $(X,Y)$,  $(X^\prime,Y^\prime)$ and $(X^{\prime\prime},Y^{\prime\prime})$ be {\it i.i.d.} copies, then 
$$
dcov^2(X,Y) = \mathbb{E}(|X-X^\prime| |Y-Y^\prime|) + \mathbb{E}(|X-X^\prime|) \mathbb{E}(|Y-Y^\prime|) - 2 \mathbb{E}(|X-X^\prime| |Y-Y^{\prime\prime}|) \;.
$$
Thus,  DCC is the correlation between the dot products which the "double centered" (it is the operation of converting the distances to the scalar products while placing the origin at the geometric center) matrices are comprised of.  It is important to mention that the definition of distance correlation coefficient can be extended to variables with finite first moments only and lack of DCC defines independence.

As both PCC and DCC quantify strength of dependence,   is important to understand how large the differences between these two measures can possibly be.  A natural question is how large the DCC can be for variables for which PCC is zero,  since uncorrelatedness only means the lack of linear dependence.  Importantly,  nonlinear or nonmonotone dependence may exist.  The fact that PCC requires finite second moments while DCC requires finite first moments implies that PCC is more sensitive to the tails of the distribution.  Although methods based on ranks (Spearman rank correlation) can be applied in some problems,  these methods are effective only for testing linear or monotone types of dependence.  Importantly,  uncorrelatedness (PCC = 0) is too weak to imply a central limit theorem which requires independence (DCC = 0) necessarily \cite{Brad-81,  Brad-88,  Brad-07,  Sz-08}.

We have used stocks listed under S\&P 500 for the time period August 2000 to August 2022 and focus on financial market crisis that occurred in the years 2008 (subprime mortgage crisis),  2010 (European debt crisis),  2011 (August 2011 stock market fall),  2015 (Great fall of China), 2020 (COVID-19 recession) and 2022 (ongoing Russo-Ukrainian war) along with bubble periods of 2002 (stock market downturn of 2002) and 2007 (Chinese stock bubble).  In order to point out the differences of using DCC,  we will also focus on epochs for which PCC $\approx  0$. 

We analyze the Pearson and Distance correlation matrices and their moments along with eigenvalue distributions and participation ratios distribution.  Participation ratios quantify the number of components that participate significantly in each eigenvector \cite{Gu-98, Pl-99}.  We show that there are strong correlations in all these three measures at the times of crisis.  Using correlation matrices to represent market states \cite{Mu-12, Gu-15, NJP-18},  we employ agglomerative clustering \cite{Plos-15} to identify correlation matrices that act similarly and compare the clustering results for the selected stocks using PCC and DCC.  

\section{Data set}

We use the 5552 daily closing prices of $N = 370$ stocks listed under S\&P 500 for the time period August 2000 to August 2022 downloaded freely from Yahoo finance webpage \cite{YF}.  The selected stocks are the ones that have been continuously traded for the chosen time period.  Using the daily closing prices $P_i(t)$,  with index $i$ representing a given stock and time $t = 1,  2, \ldots, T$.  daily returns $r_i(t) = [P_i(t) - P_{i-1}(t)]/P_{i-1}(t)$ are calculated.  Here $T$  is total number of the trading days present in the considered time horizon.  We then have 5551 daily returns and use these to compute the equal-time cross-correlation matrices based on PCC and DCC. Table \ref{t-sect} gives the distribution of the sectors.

\begin{table}
  \centering
    \begin{tabular}{|l|c|c|c|}
    \hline
    \textbf{Sector} & \textbf{Ticker} & \textbf{Stocks} & \textbf{Weight} \\
    \hline
    {Communication Services} & {TS} & 11  & 0.03 \\
    Consumer Discretionary & CD & 38 & 0.10 \\
    Consumer Staples & {CS} & 27  &  0.07 \\
    {Energy} & {EN} & 18    &  0.05 \\
    {Financials} & {FI} & 49 &  0.13 \\
    Health Care & HC & 51    & 0.14  \\
    Industrials & {IN} & 54    &  0.15 \\
    Information Technology & {IT} & 49 & 0.13 \\
    Materials & {MA} & 21    & 0.06 \\
    {Real Estate} & {RE} & 25 & 0.07 \\
    {Utilities} & {UT} & 27  & 0.07 \\
    \hline    \hline
    \end{tabular}
    \caption{Distribution of the constituent sectors of selected stocks of financial market S\&P 500. }
  \label{t-sect}
\end{table}

The disadvantage of working with long financial time series is the loss of information over short periods of time,  it is convenient to divide it into short time series (epochs).  Computing returns and dealing with epochs guarantees (weakly) stationary time series.  With this,  one can study the evolution over time,  for example,  of the average correlations.  This helps focus on details in a given particular time interval as financial market is a dynamic entity.  

First, we analyze the distribution of correlation matrix elements,  eigenvalues and participation ratios obtained using both PCC and DCC for all the 138 time epochs (non-overlapping epochs of size 40 days each).  We show that there are strong correlations in all these three measures at the times of crisis.  In other words,  there is collective motion during crashes.  

\section{Correlations and spectral analysis}

\begin{figure}
\centering
    \includegraphics[width=6.75cm,height=6.75cm]{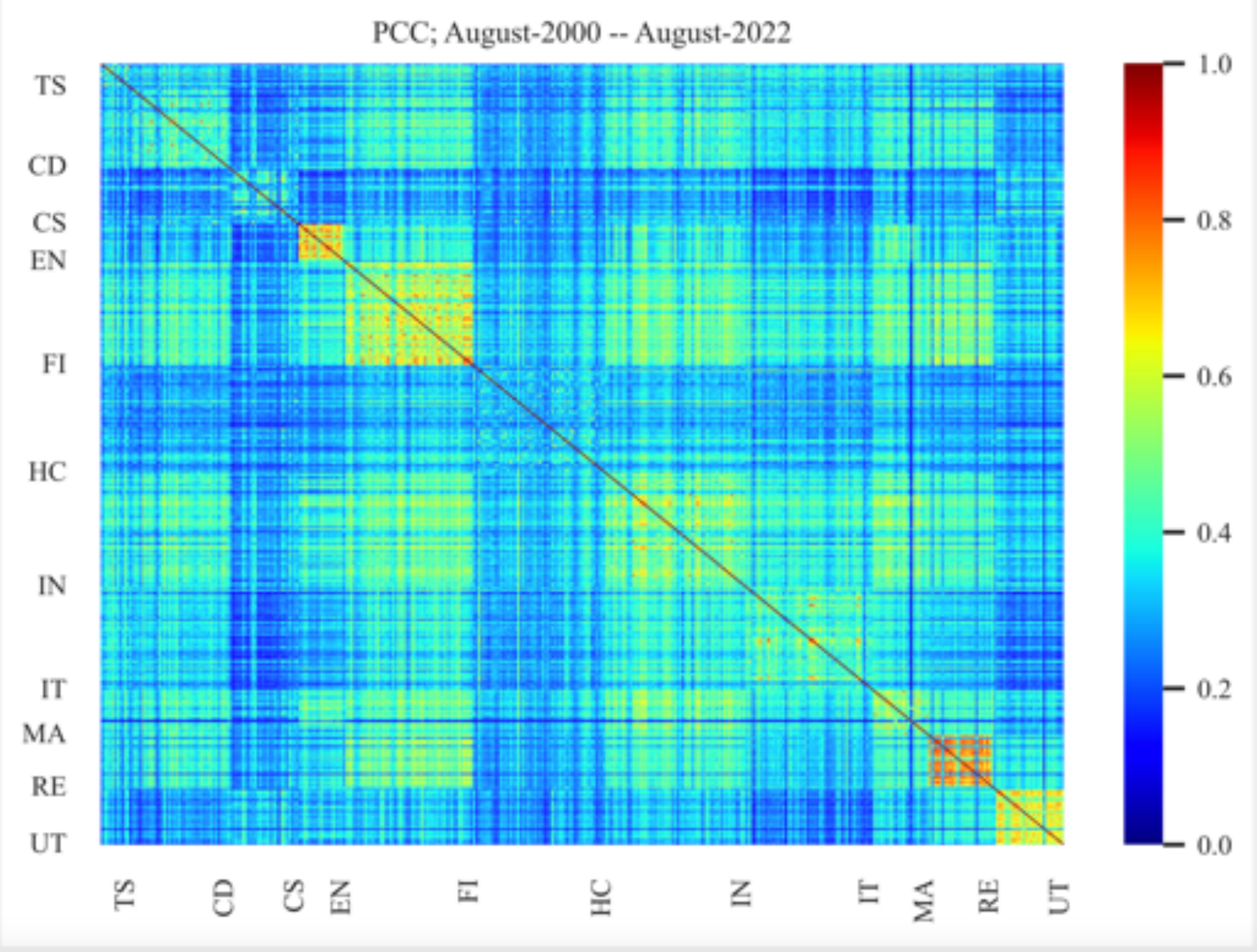}
    \includegraphics[width=6.75cm,height=6.75cm]{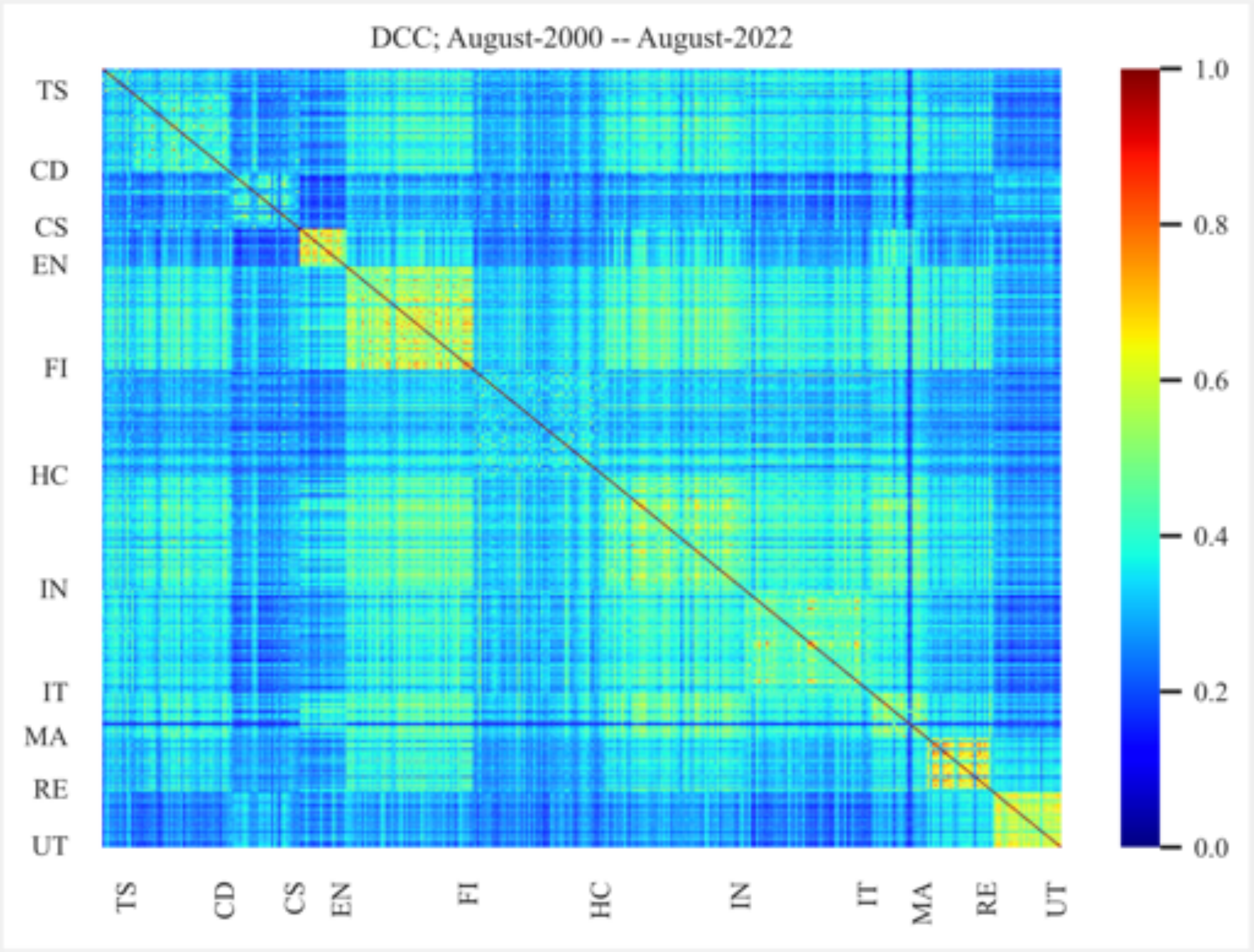}
    \caption{Correlation matrices for the total time horizon considered.  Left panel shows the PCC matrix and the right panel shows the DCC matrix.  The minimum values for PCC and DCC are 0.003 and 0.06 respectively.  Similarly,  the average PCC and DCC are 0.347 and 0.34.}
    \label{fig-1}
\end{figure}

To begin with,  we plot the correlation matrices obtained using PCC and DCC in Fig.  \ref{fig-1}.  As expected,  one loses the details due to long time averaging.  We plot both PCC and DCC correlation matrices on the same scale as there are no negative correlations in the PCC matrix computed for the total time horizon.  Sectorial correlations are stronger for PCC in comparison to DCC.

\begin{figure}
\centering
    \includegraphics[width=6.75cm,height=6.75cm]{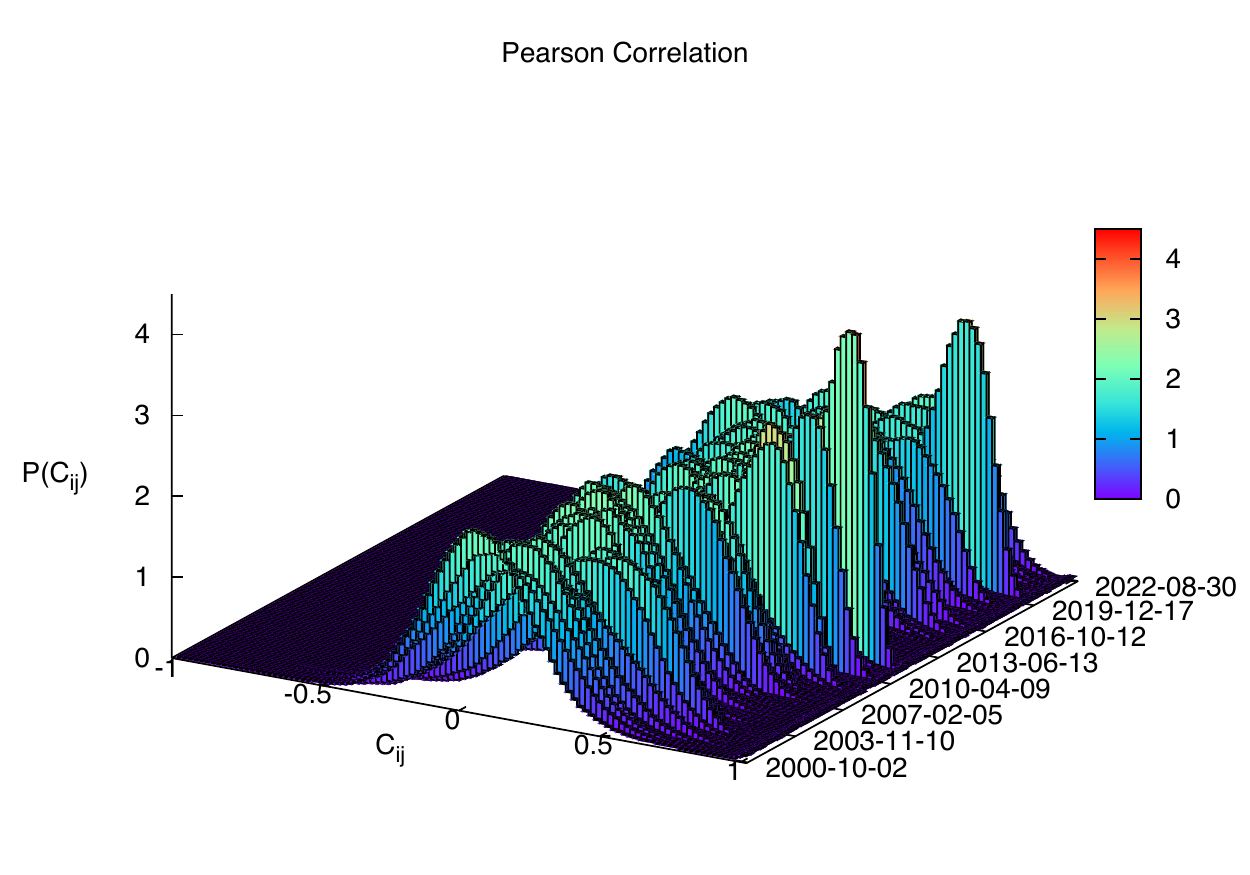}
    \includegraphics[width=6.75cm,height=6.75cm]{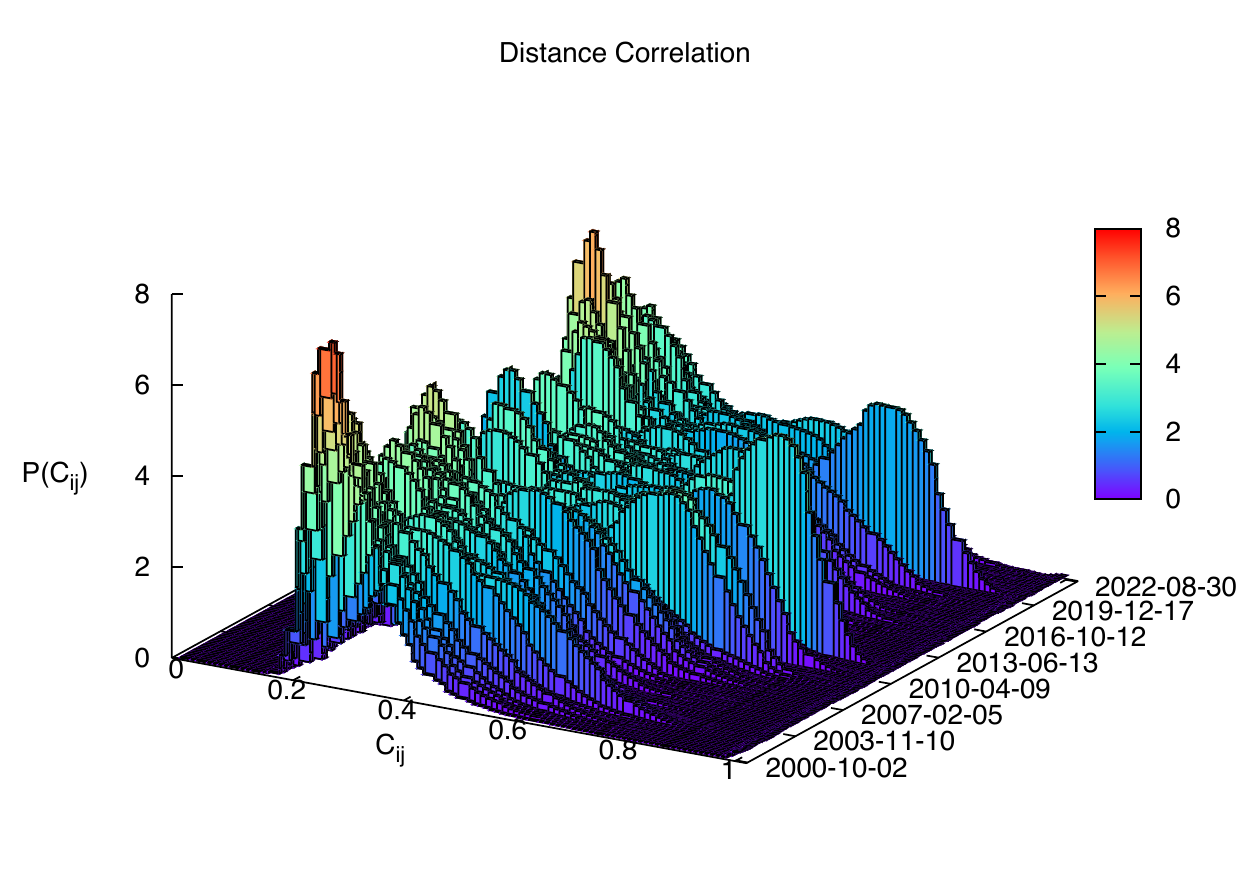} \\
   \includegraphics[width=7.75cm,height=4.75cm]{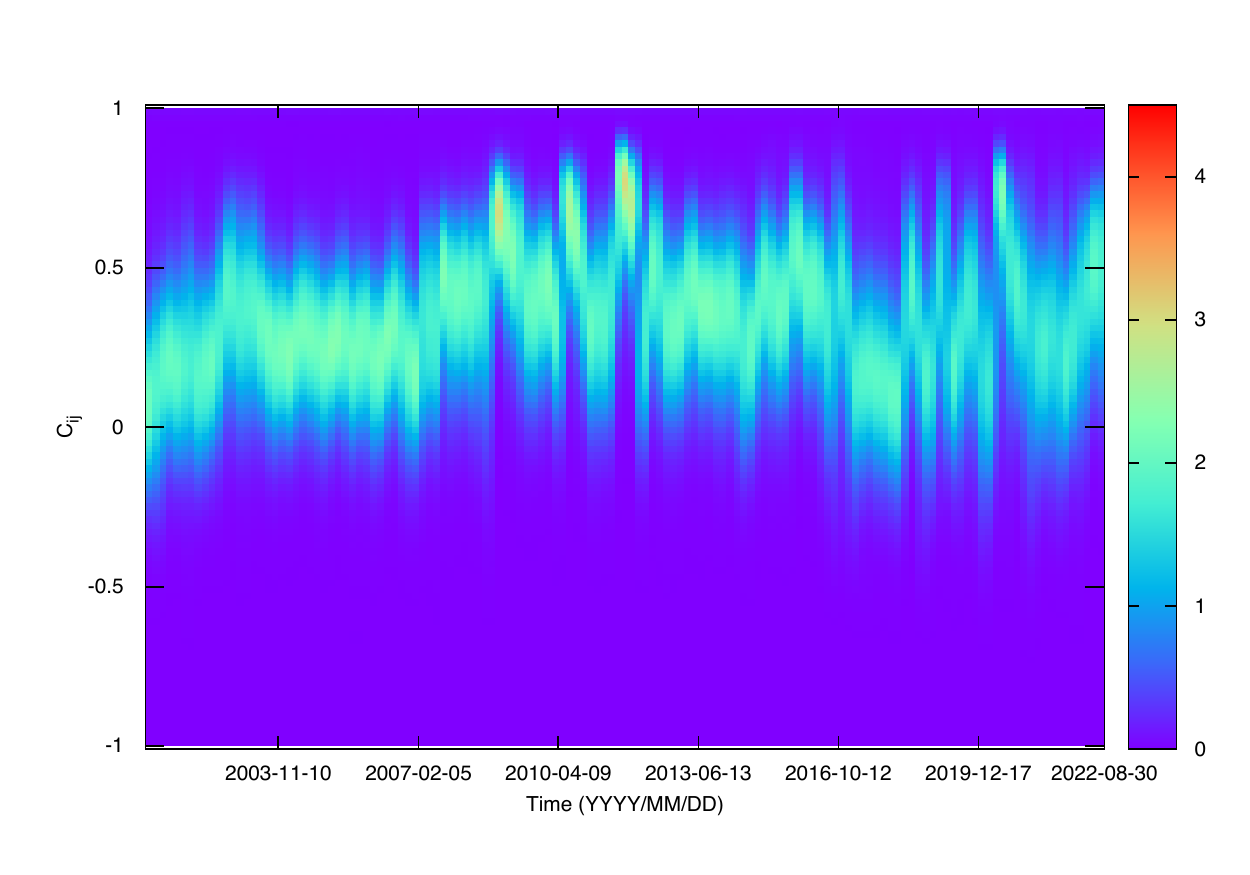}
    \includegraphics[width=7.75cm,height=4.75cm]{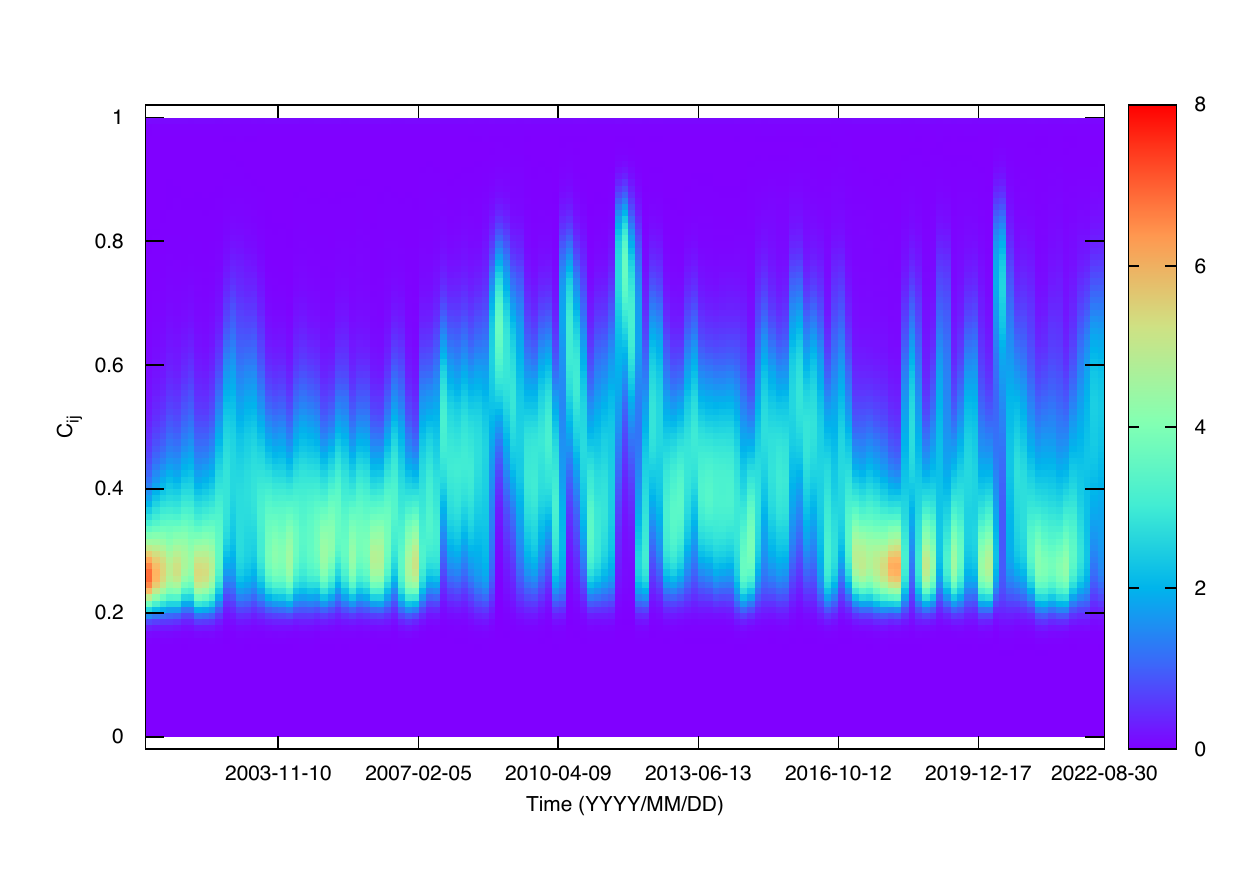}
    \caption{Time evolution of the distribution of correlation matrix elements $P(C_{ij})$.  Left panel shows the $P(C_{ij})$ for PCC and the right panel shows $P(C_{ij})$ for the DCC.  Bottom panel shows the 2D projection of the corresponding figures in the top panel. }
    \label{fig-2}
\end{figure}

As one can not see any specific structures in the plots for the correlation matrices for the complete time horizon, we study the distribution of correlation matrix elements for each epoch as shown in Fig. \ref{fig-2}.  DCC and PCC both show a clear shift towards higher values of correlation during the crisis periods of interest (2002, 2008, 2010, 2011, 2020 and 2022).  Also,  DCC shows the peaks of distributions at lower values of correlation for the non-crisis periods, unlike PCC.  Notably, DCC $\geq 0.2$ for the time horizon considered implying that there are non-monotonic correlations present in financial markets at all times.

\begin{figure}
\centering
    \includegraphics[width=7.75cm,height=4.75cm]{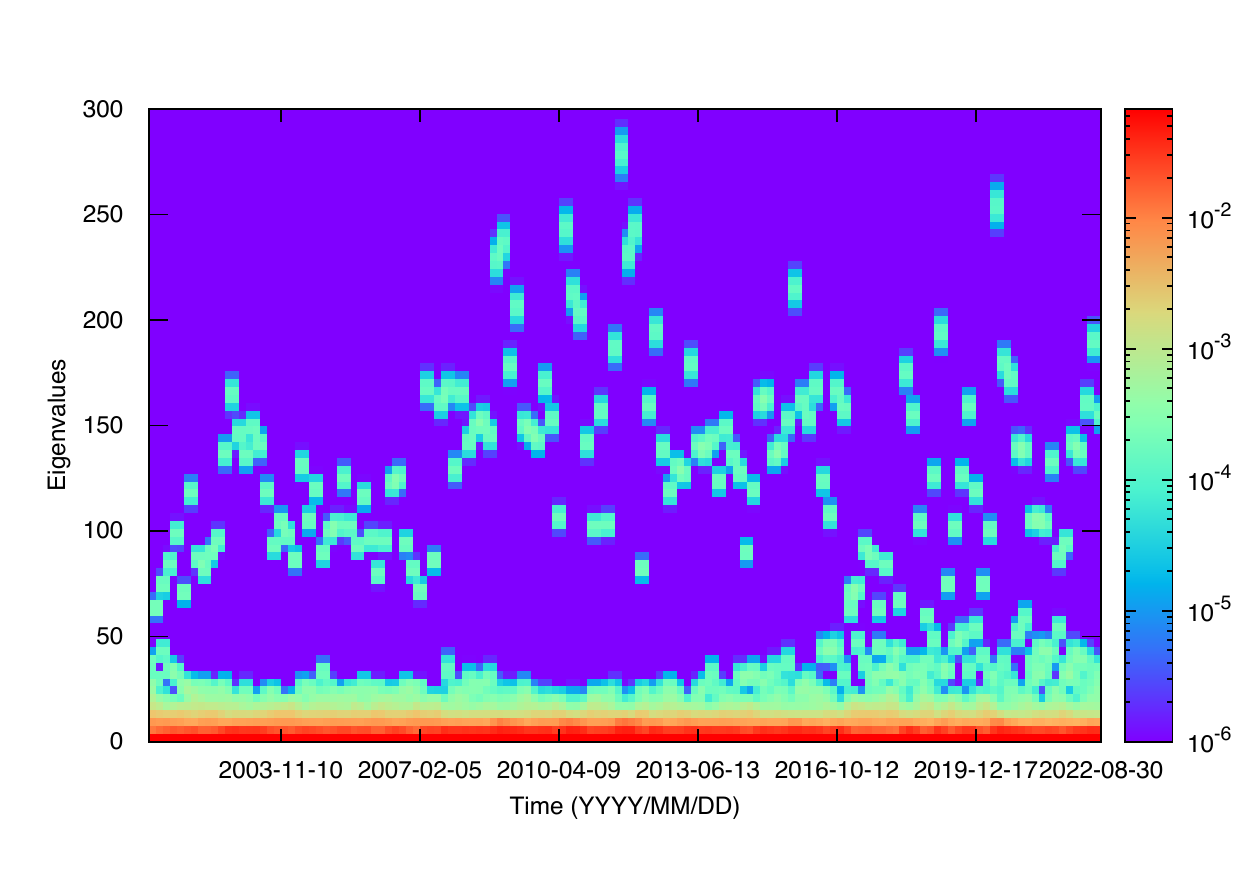}
    \includegraphics[width=7.75cm,height=4.75cm]{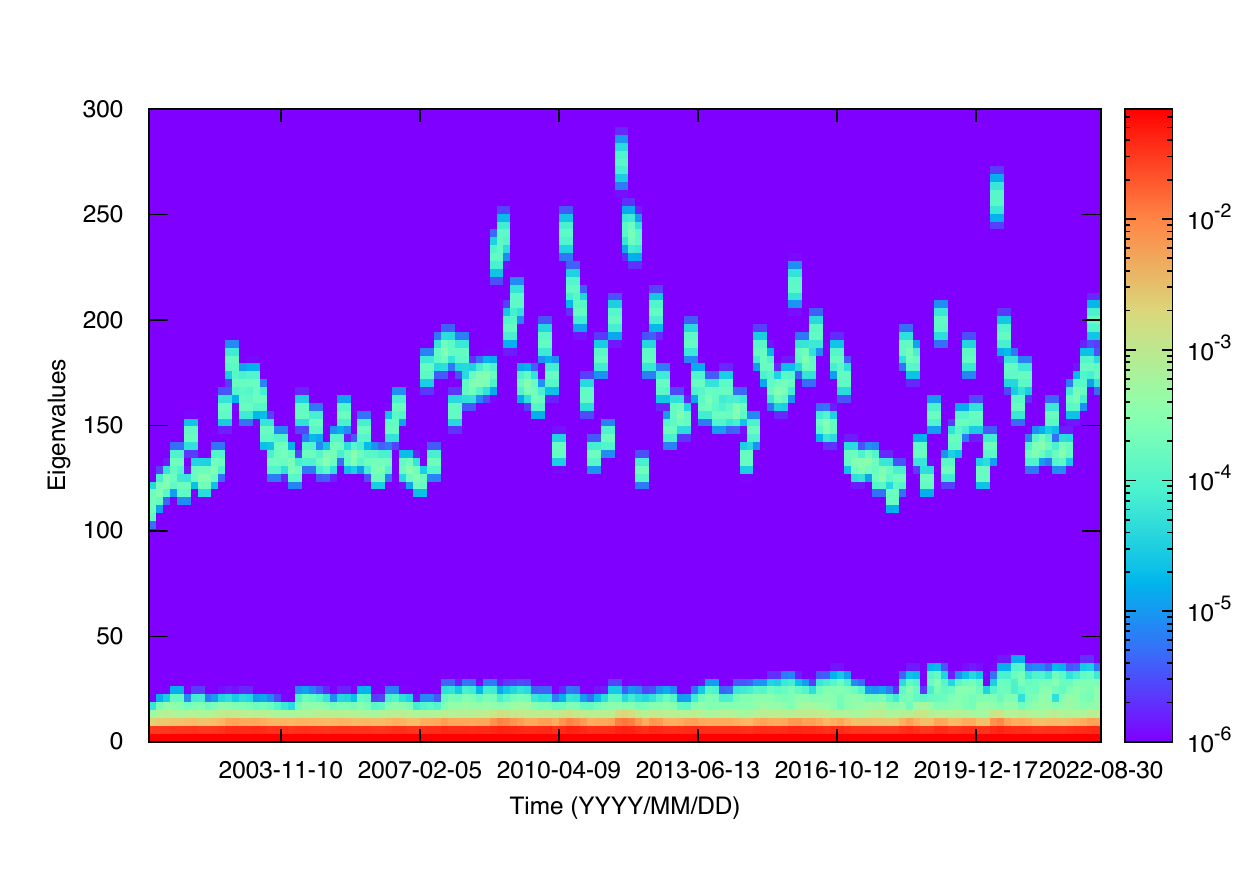} \
    \caption{Time evolution of the distribution of eigenvalues.  Left panel shows the eigenvalue distribution for PCC and the right panel shows for the DCC.  }
    \label{fig-3}
\end{figure}

Next, we analyze the time evolution of distribution of eigenvalues of correlation matrices as shown in Fig. \ref{fig-3}; note that the plot is logarithmic.  All the correlation matrices are singular and thus, we have a delta peak at zero eigenvalues in addition to bulk distribution (which follows random matrix theory predictions) and outliers that  represent correlations \cite{Ed-88, La-99, Pl-99, Ma-18, Deo-19, Pha-RMT}.  The largest eigenvalue, which is linearly correlated with average correlations,  attains very large values in crisis periods as seen from distribution of eigenvalues for both PCC and DCC.  Around end of 2016, the gap between the bulk eigenvalue distribution and outliers for PCC is little.  One can clearly see a comparatively broad bulk eigenvalue distribution for DCC beyond 2019.  This feature is also seen in the plot for PCC, however it is equally broad for 2001 when the largest eigenvalue is $< 100$.  Like the distribution of correlation matrix elements in Fig. \ref{fig-1}, eigenvalue distributions for DCC and PCC both show a clear shift towards higher values of outliers during the crisis periods of interest (2002, 2008, 2010, 2011, 2015, 2020 and 2022). 

\begin{figure}
\centering
    \includegraphics[width=7.75cm,height=4.75cm]{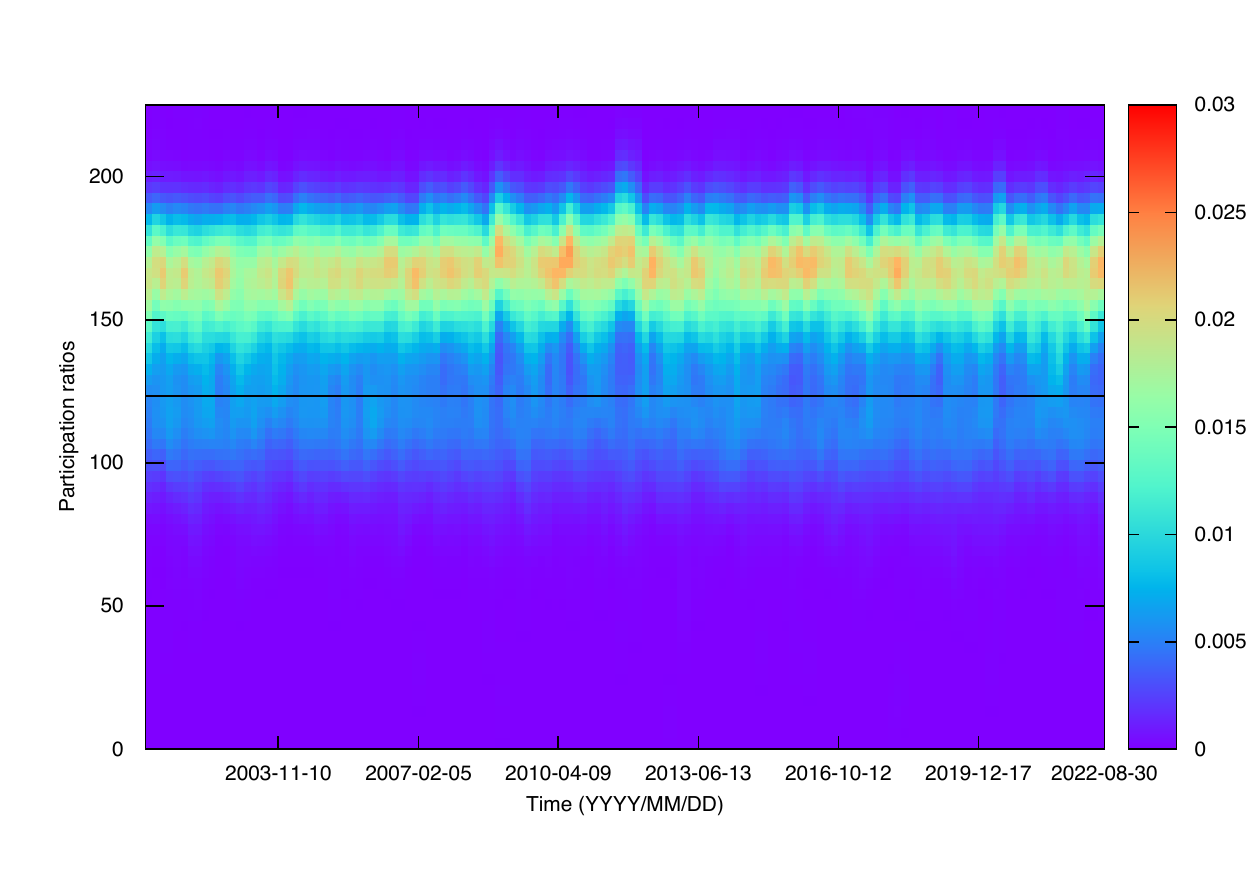}
    \includegraphics[width=7.75cm,height=4.75cm]{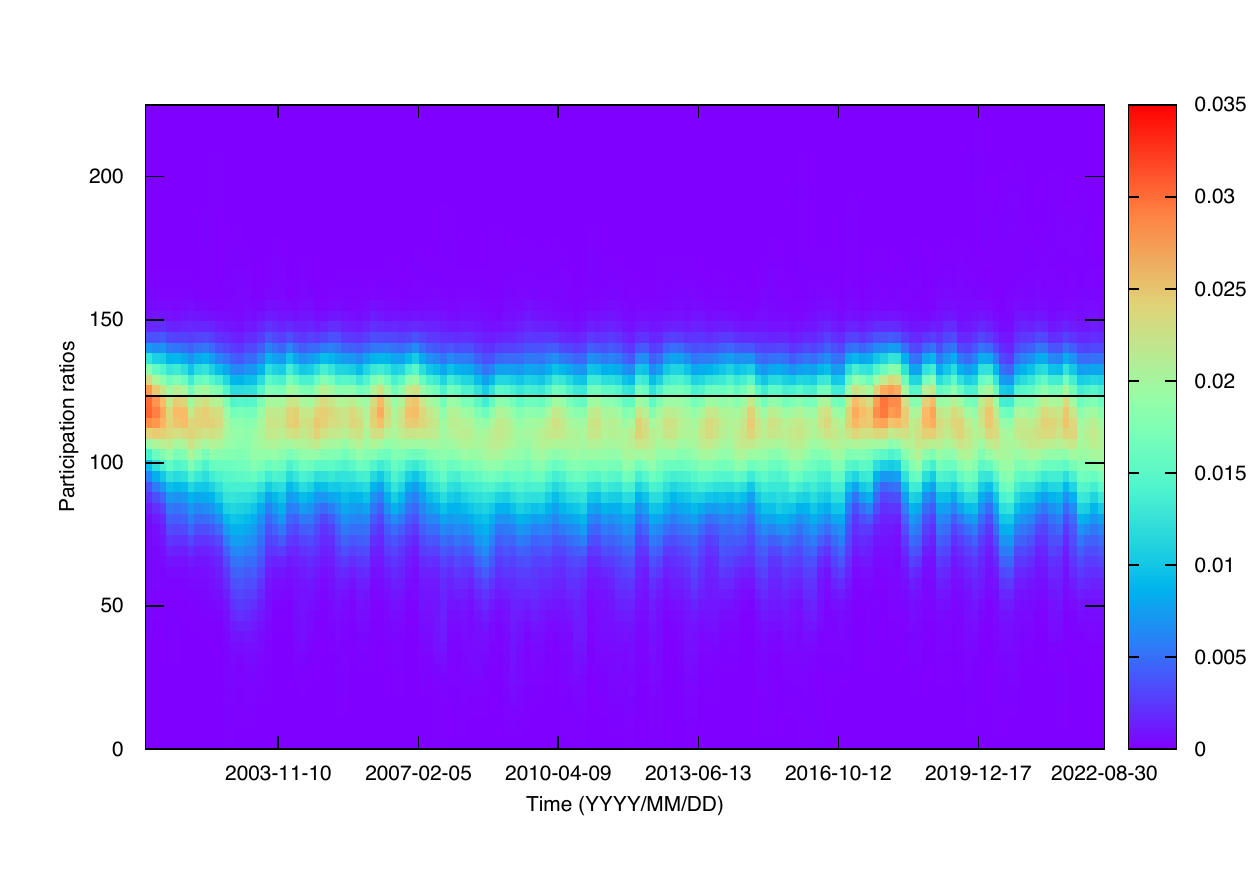} \
    \caption{Time evolution of the distribution of participation ratios.  Left panel shows the participation ratios distribution for PCC and the right panel shows for the DCC.  The horizontal line shows the expectation value obtained from random matrix theory.}
    \label{fig-4}
\end{figure}

We use participation ratio (PR) to quantify the number of components that participate significantly in each eigenvector $\nu_i$,
\begin{equation}
PR_\nu = \left[{\displaystyle\sum_{i=1}^N |\nu_i|^4 }\right]^{-1} \;.
\label{eq-1}
\end{equation}
PR gives the number of elements of an eigenvector that are different from zero that contribute significantly to the value of the eigenvector and thus, takes values between 1 (only one component) and $N$ (all components contributing equally).  The expectation value of PR for a Gaussian Orthogonal Ensemble (classical random matrix ensemble) has the limiting value of $\langle PR \rangle \approx N/3$ \cite{Gu-98, Ko-book}.  We show the time evolution of distribution of PR for PCC and DCC in Fig. \ref{fig-4}.  The horizontal line in the plots gives the average PR value estimated using Gaussian Orthogonal Ensemble.  As seen from the plots, the average PR for PCC is $\approx 160$ while that for DCC is $\approx 110$.  The distribution of PR in case of PCC shows a slight upward shift during crisis years of 2008, 2010 and 2011 while we see a slight downward shift in case of DCC during the crisis years 2002, 2008, 2010, 2011 and 2020.  The lesser the average correlation, prominent is the downward shift in the distribution of PR in case of DCC. 

\begin{figure}
\centering
    \includegraphics[width=12.75cm,height=10.75cm]{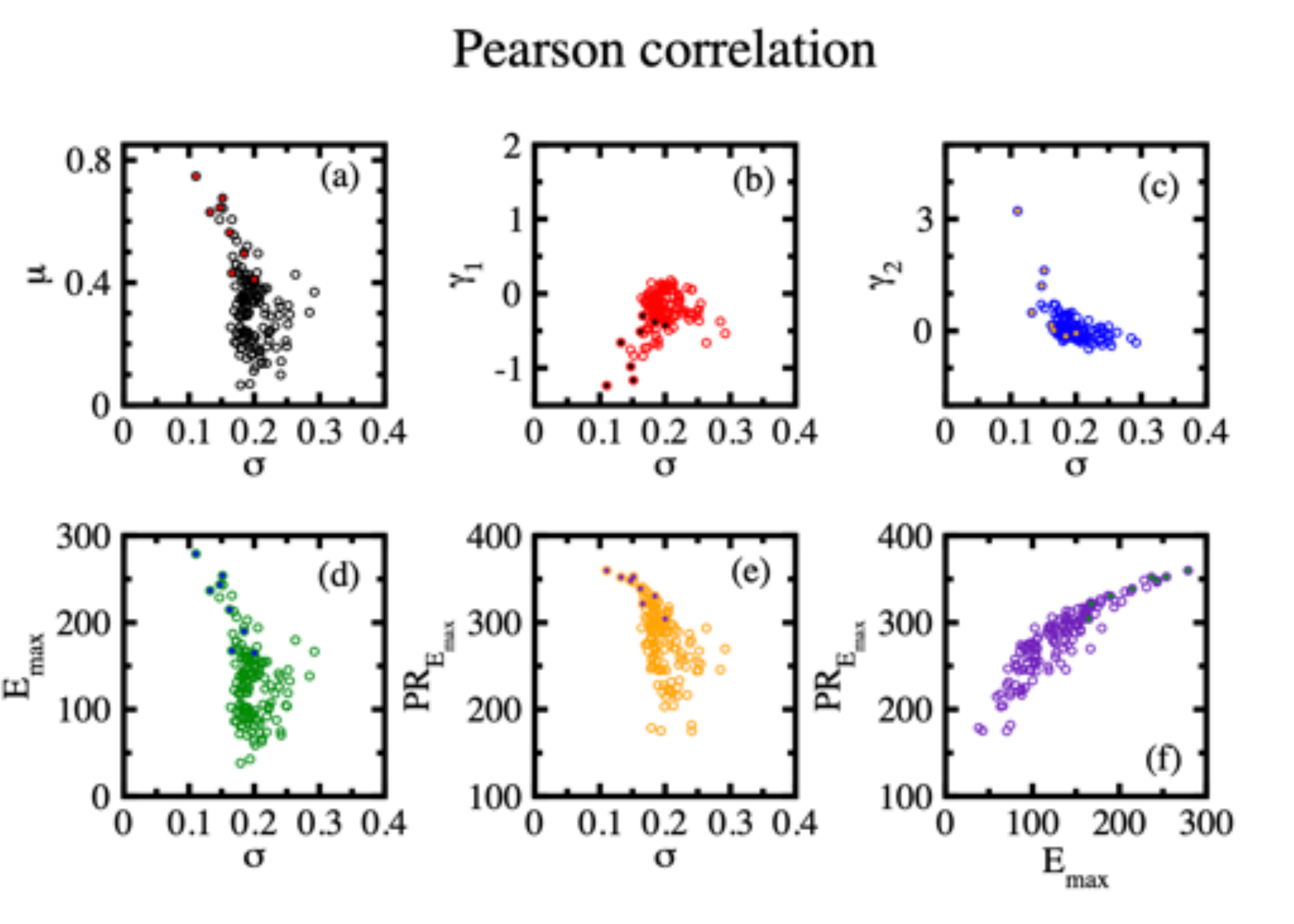} 
    \includegraphics[width=12.75cm,height=10.75cm]{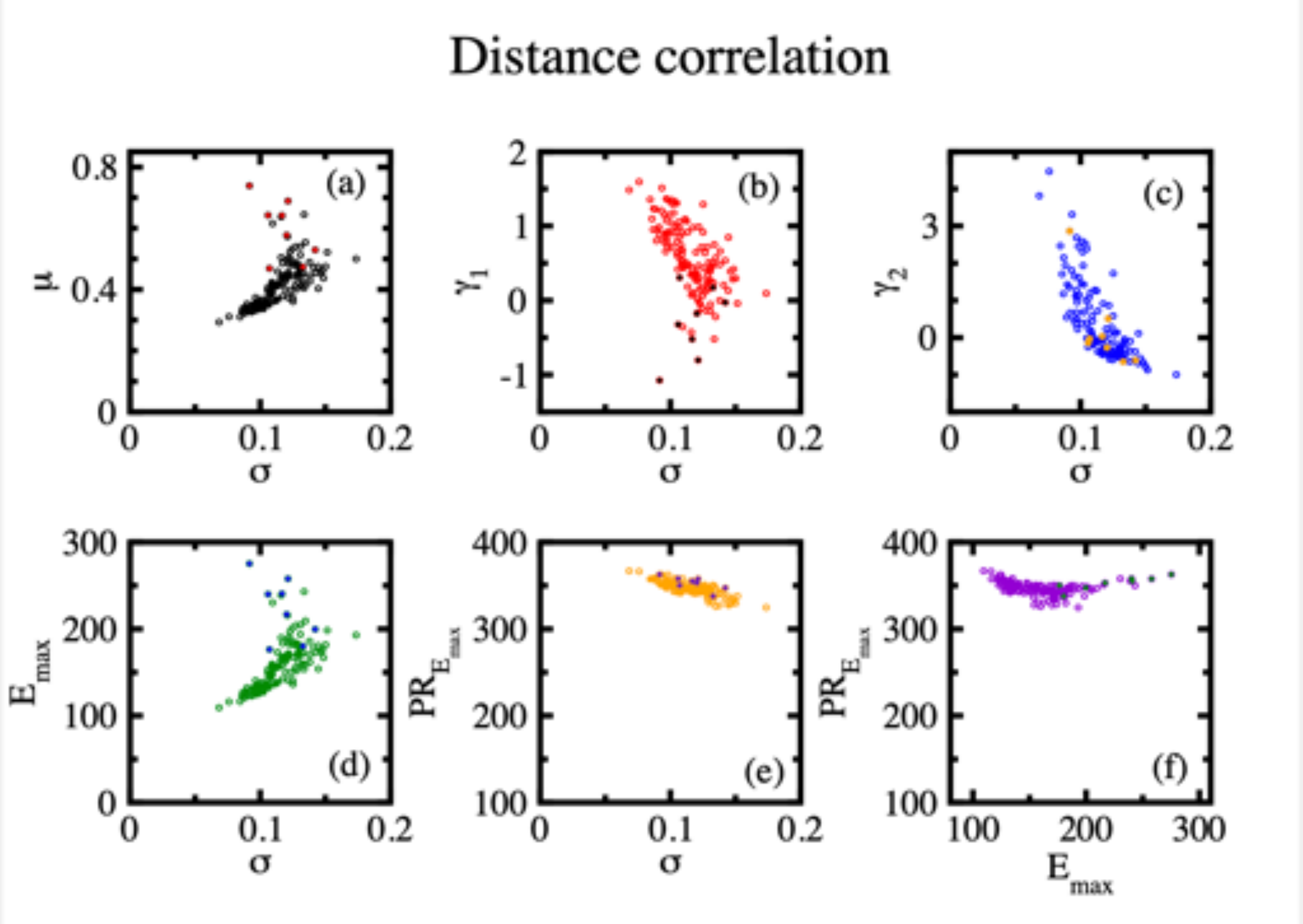} 
    \caption{Scatter plots corresponding to PCC (top panel) and DCC (bottom panel) between (a) mean correlation $\mu$ and standard deviation $\sigma$, (b) skewness $\gamma_1$ and $\sigma$,  (c) excess kurtosis $\gamma_2$ and $\sigma$,  (d) largest eigenvalue $E_{max}$ and $\sigma$, (e) PR for the largest eigenvalues $PR_{E_{max}}$ and $\sigma$, and (f) PR for the largest eigenvalue $PR_{E_{max}}$ and largest eigenvalues $E_{max}$. }
    \label{fig-6}
\end{figure}

Next we analyze the scatter plots between various moments \cite{St-87} corresponding to PCC and DCC and the results are presented in Fig. \ref{fig-6}.  Note that each point corresponds to an epoch and we represent the bubble and crisis periods of interest as solid circles. As seen in Fig. \ref{fig-1},  the crisis periods appear at higher values of mean correlations $\mu$ for both PCC and DCC.  For PCC, the crisis periods of 2008, 2010, 2011, 2015 and 2020 appear with largest $\mu$ while the bubble periods of 2002 and 2007 alongwith the ongoing Russo-Ukrainian war have relatively lower values of $\mu$.  Skewness is negative for all the crisis periods and the bubble periods implying that the distribution has a longer left tail and bulk is concentrated towards the right side.  Kurtosis for the crisis periods of 2008, 2010, 2011, 2015 and 2020 is positive implying the distributions are leptokurtic while distributions are platykurtic for the bubble periods of 2002 and 2007,  and the ongoing Russo-Ukrainian war.  $E_{max}$ reflects a similar behavior as average correlations $\mu$ and $PR_{E_{max}}$ is also maximum for crisis periods of 2008, 2010, 2011, 2015 and 2020.  In summary, PCC distinguishes the bubble periods of 2002 and 2007, and the ongoing Russo-Ukrainian war from the crisis periods of 2008, 2010, 2011, 2015 and 2020 depending on kurtosis of the distribution of correlation matrix elements.

Similarly, in case of DCC: for $\mu < 0.5$, $\sigma$ increases with increasing $\mu$ and for $\mu > 0.5$, $\sigma$ decreases with increasing $\mu$.  The crisis periods of 2008, 2010, 2011, 2015 and 2020 appear with largest $\mu$ while the bubble periods of 2002 and 2007, and the ongoing Russo-Ukrainian war have relatively lower values of $\mu$.  Skewness is negative for the crisis periods of 2008, 2010, 2011, 2015 and 2020 implying that the distribution has a longer left tail and bulk is concentrated towards the right side, while distribution has a longer right tail for the bubble periods of 2002 and 2007, and distribution is symmetric for the ongoing Russo-Ukrainian war.  Kurtosis for the crisis periods of 2010, 2011 and 2020 is positive implying the distributions are leptokurtic while distributions are platykurtic for the bubble periods of 2002 and 2007, crisis periods of 2008 and 2015, and the ongoing Russo-Ukrainian war.  $E_{max}$ reflects a similar behavior as average correlations $\mu$ and $PR_{E_{max}}$ is constant around the maximum value for all the epochs.  In summary,  DCC distinguishes the bubble periods from the crisis periods depending on skewness of the distribution of correlation matrix elements.  Also,  DCC distinguishes the bubble periods of 2002 and 2007,  the crisis periods of 2008 and 2015, and the ongoing Russo-Ukrainian war from the crisis periods of 2010, 2011 and 2020 depending on kurtosis of the distribution of correlation matrix elements.

\section{Agglomerative clustering}

In this section, we compare the clustering results for the selected stocks using PCC and DCC.  We employ agglomerative clustering that creates clusters by successively merging epochs starting with singleton clusters. Using the linkage criterion in each iteration, the clusters are joined together until obtaining a single cluster \cite{Plos-15}.  Dendrograms give the representation of this hierarchy.  Choosing the threshold value then decides the number of clusters that will be obtained.  We cluster similar correlation matrices into these optimized $n$ number of “market states”.  This is a variance-minimizing approach tackled with an agglomerative hierarchical approach.  Dendrograms obtained for the PCC and DCC are given in Appendix \ref{app-1}.  

\begin{figure}
\centering
    \includegraphics[width=7.75cm,height=6.75cm]{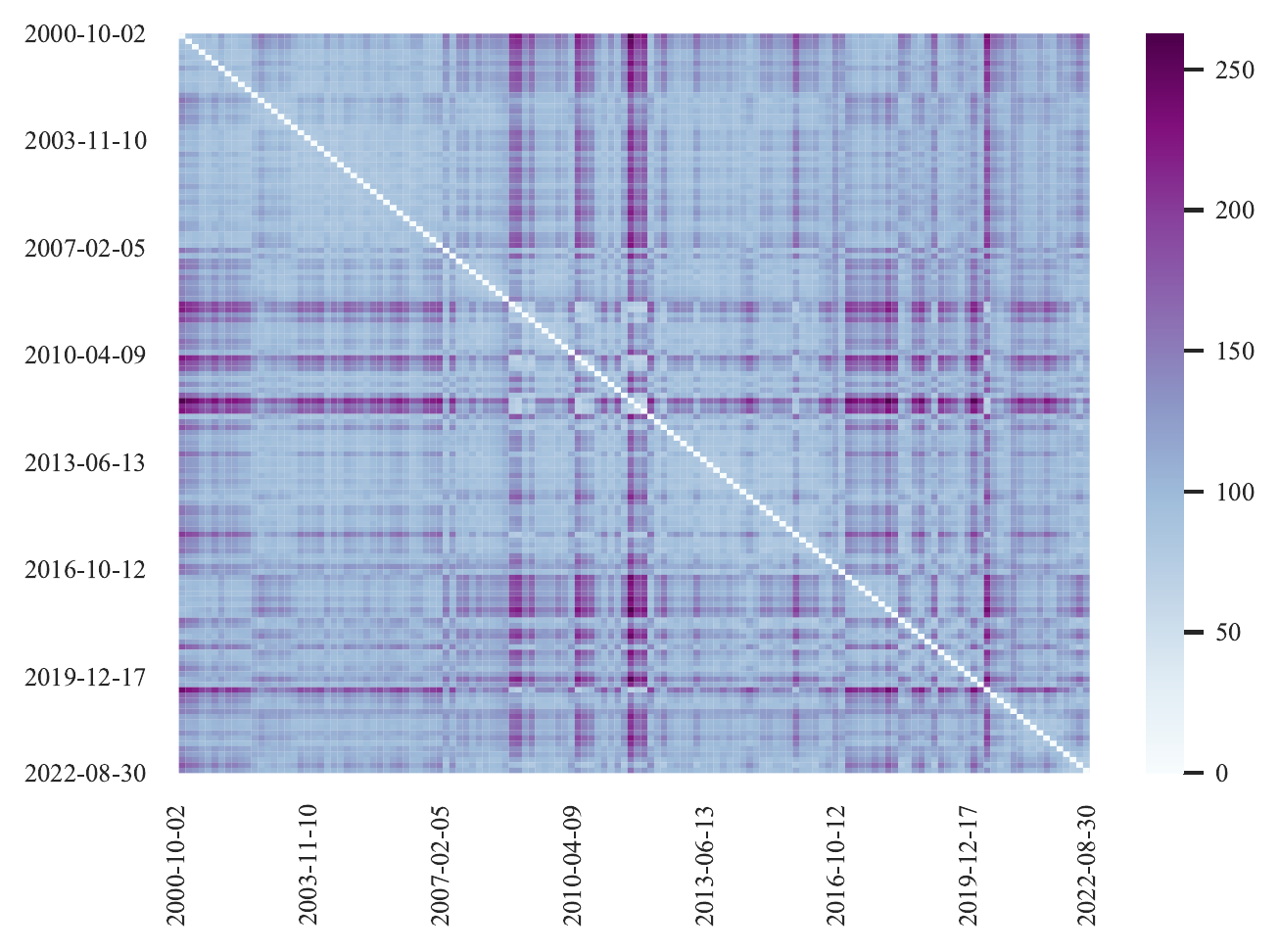} 
    \includegraphics[width=7.75cm,height=6.75cm]{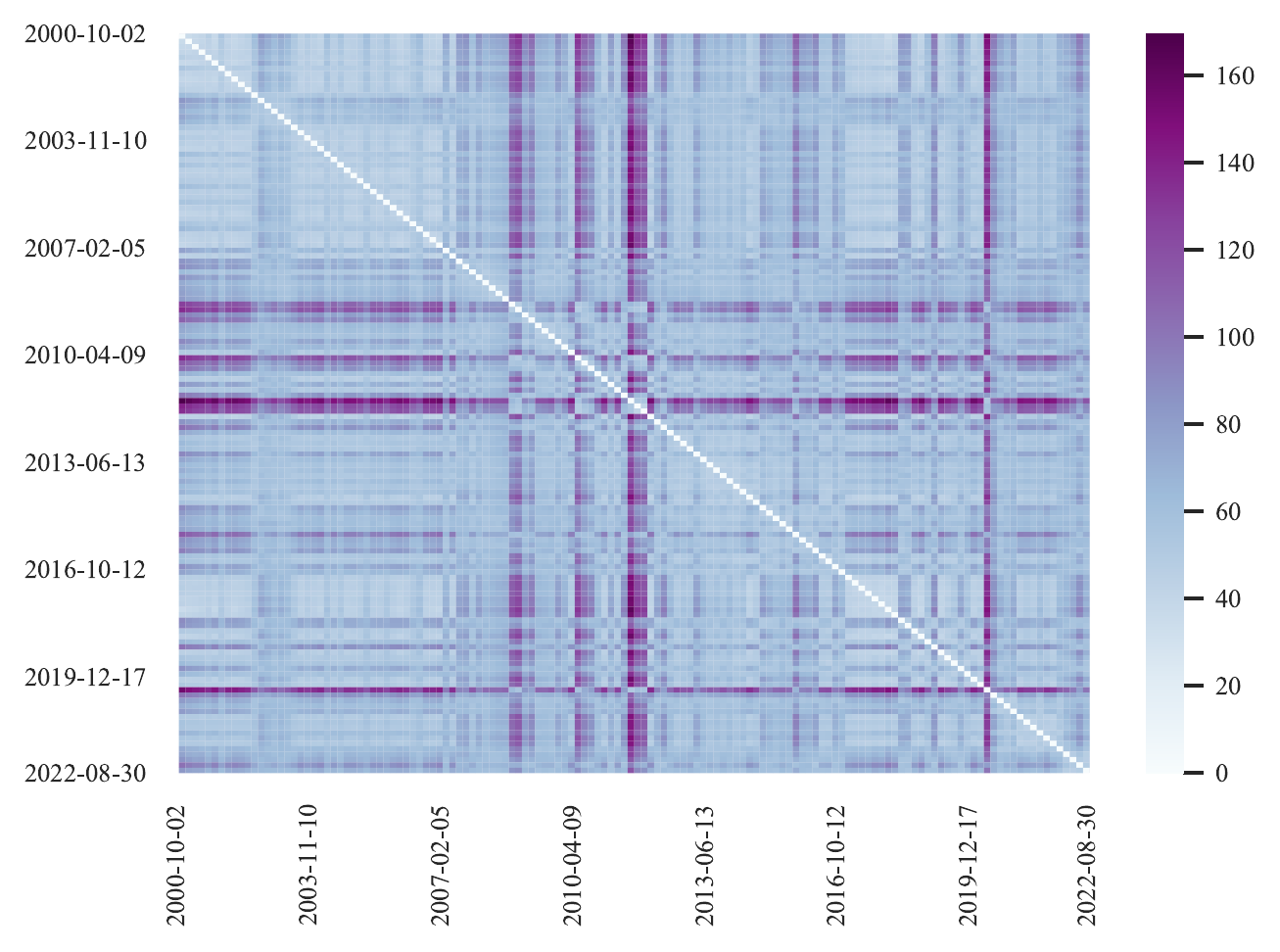} 
    \caption{Euclidean distance matrix obtained using Eq. \eqref{eq-2} for PCC (left panel) and DCC (right panel).}
    \label{fig-7}
\end{figure}

\begin{figure}[h]
\centering
    \includegraphics[width=14.75cm,height=10.75cm]{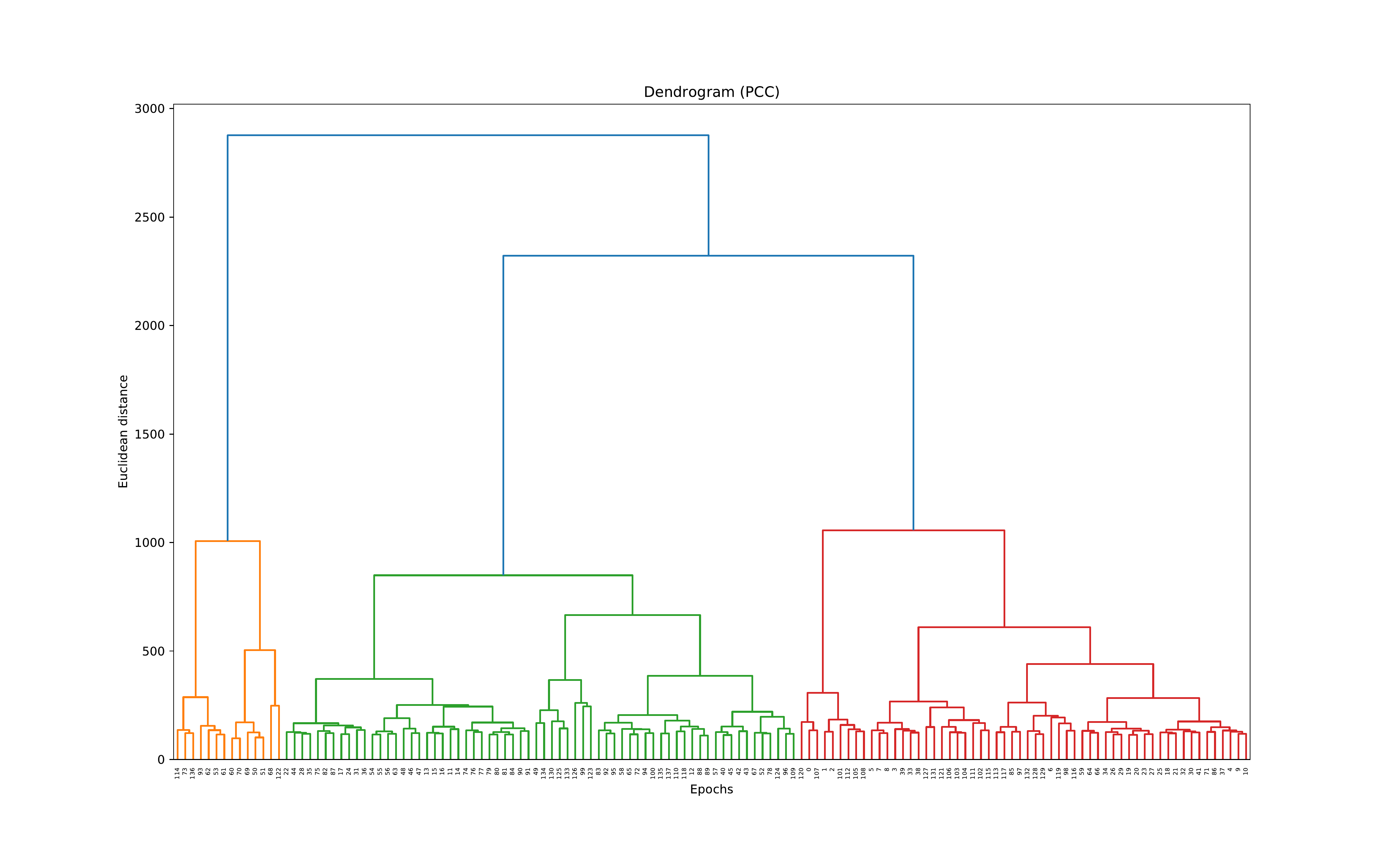} 
    \caption{Dendrogram obtained for PCC using agglomerative clustering.}
    \label{fig-8a}
\end{figure}

\begin{figure}[h]
\centering
    \includegraphics[width=14.75cm,height=10.75cm]{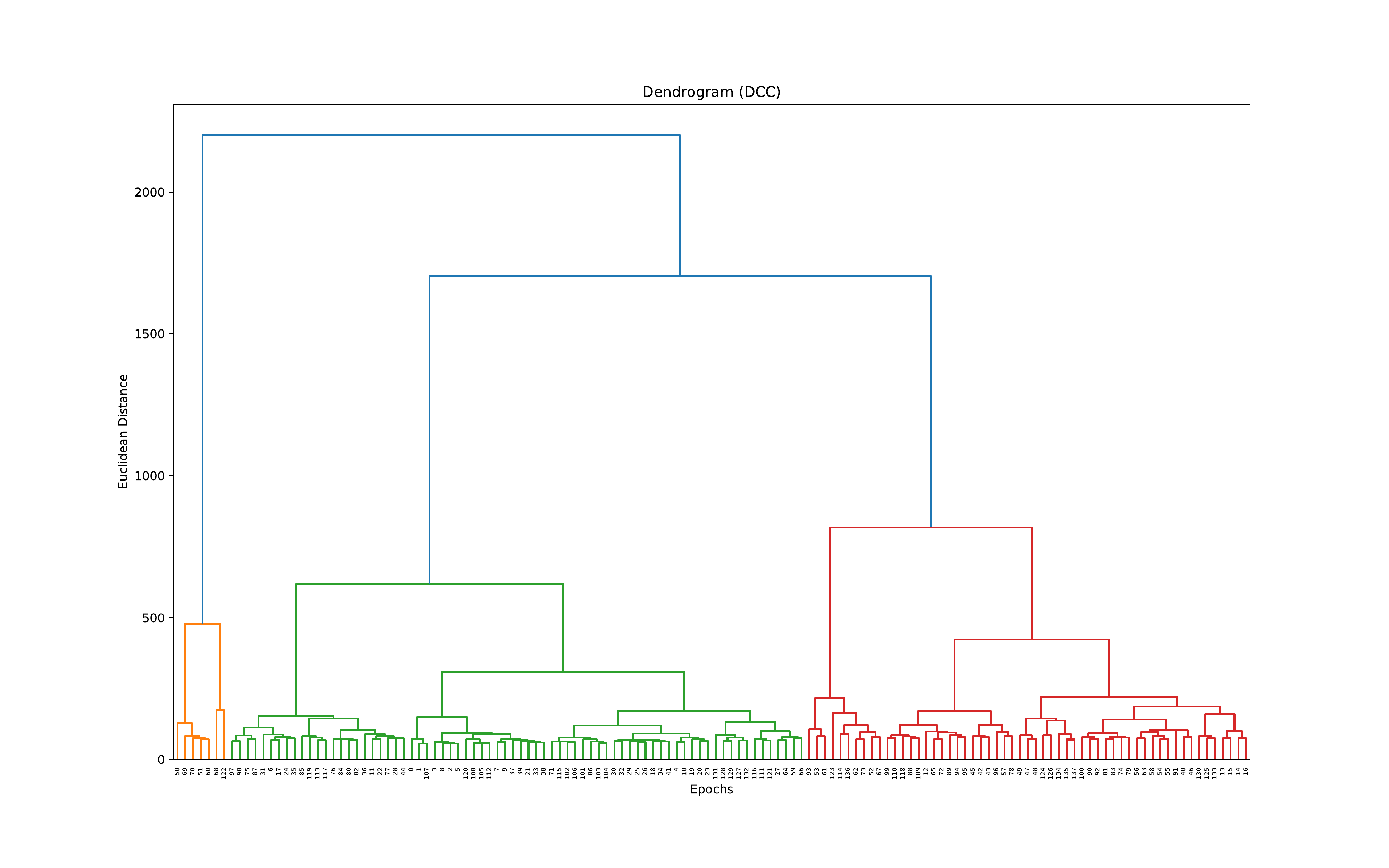} 
    \caption{Dendrogram obtained for DCC using agglomerative clustering.}
    \label{fig-8b}
\end{figure}

In order to implement this algorithm, we need to compute the distance matrix $\xi$ based on correlation coefficients $C$'s,
\begin{equation}
\xi(t_i,t_j) = d_E |C(t_i) - C(t_j)| \;,
\label{eq-2}
\end{equation}
with $d_E$ representing the Euclidean norm and indices $i,j = 1,2,,\ldots,138$ representing different epochs.  Figure \ref{fig-7} gives the Euclidean matrices for PCC and DCC respectively.  Note that the crash periods of 2008, 2010, 2011 and 2020 are visible in these.  Once the algorithm was trained with its respective distance matrix, the average correlation coefficients PCC and DCC were used as inputs to be able to group them into $n=5$ clusters that were considered adequate; see Figs. \ref{fig-8a} and \ref{fig-8b} for corresponding dendrograms.  

\begin{figure}
\centering
    \includegraphics[width=15cm]{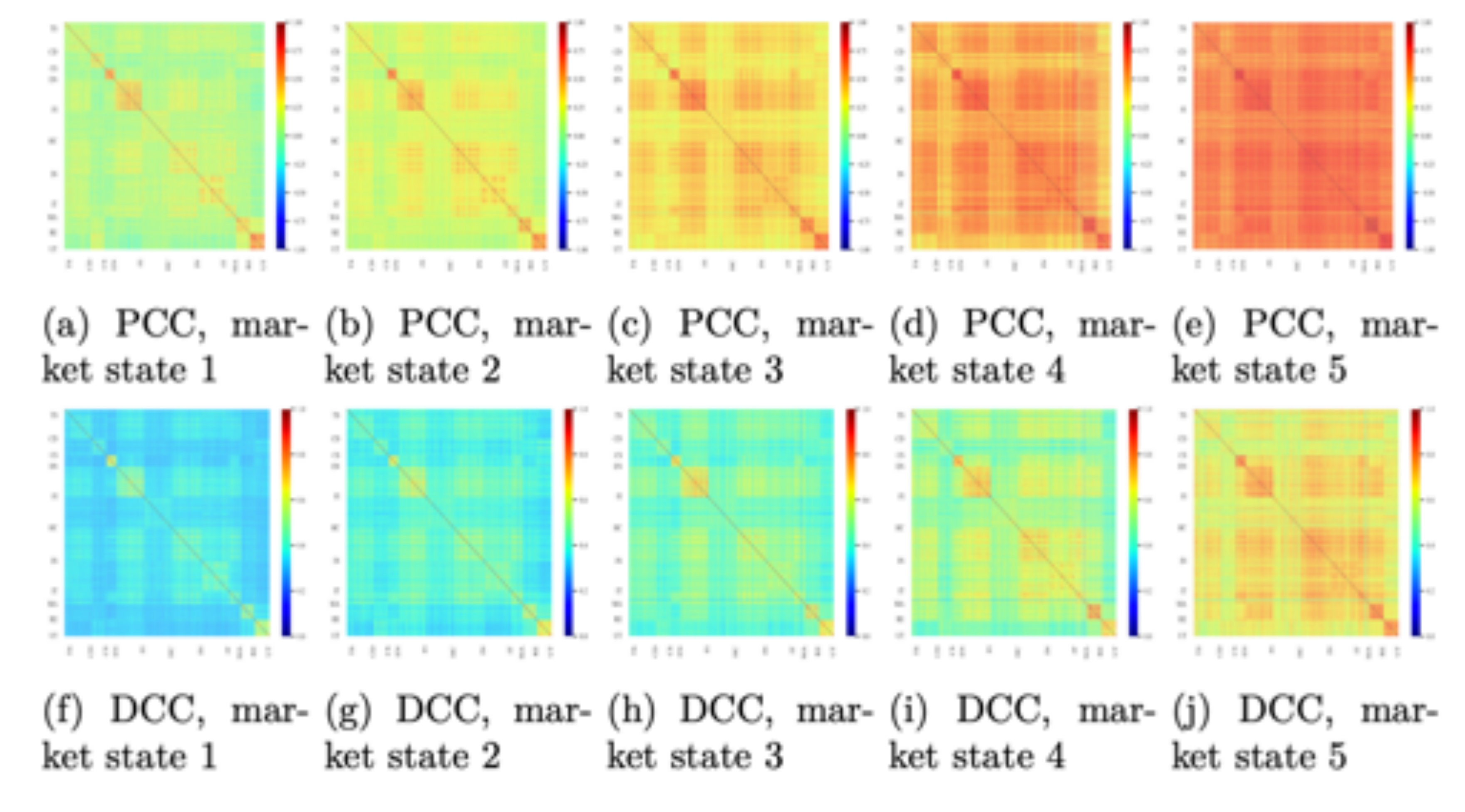} 
   \caption{Average correlation matrices for each market state obtained using agglomerative clustering for PCC [(a)-(e)] and DCC [(f)-(j)]. The average correlation coefficients (from left to right) are PCC: 0.12, 0.22, 0.37, 0.52,  and 0.65; DCC: 0.35, 0.41, 0.46, 0.54, and 0.66, respectively.}
    \label{fig-11}
\end{figure}

The average correlation matrices of each market states corresponding to both (a) PCC and (b) DCC are shown in Fig.  \ref{fig-11}.  The correlation structures vary for each market state corresponding to PCC and DCC. The average correlation coefficients (from left to right) are (a) PCC: 0.12, 0.22, 0.37, 0.52,  and 0.65, (b) DCC: 0.35, 0.41, 0.46, 0.54, and 0.66.  The number of matrices that are grouped together in each of the market states (from left to right) are (a) PCC: 9, 49, 66, 7,  and 7 and (b) DCC: 51, 23, 47, 10,  and 7.  The market states with highest correlation coefficient are 7 for both PCC and DCC. 
For PCC,  the market state with highest average correlation includes the crash periods of 2008, 2010, 2015 and 2022 with two matrices not belonging to crash periods.  For DCC, the market state with highest average correlation includes the crash periods of 2008, 2010, 2011 and 2020.  For PCC, the market state with second highest average correlation includes epochs in the vicinity of the crash periods of 2008, 2010, 2011 and 2020 and for DCC,  the market state with second highest average correlation includes epochs in the vicinity of the crash periods of 2015 and 2022.  The bubble periods of years 2002 and 2007 are included in the market state with third highest average correlation for both PCC and DCC.  There are two epochs for which PCC $\approx 0$ and these epochs are in the market state corresponding to the lowest average correlation coefficient both for PCC and DCC.  Note that this market state has respectively 9 and 51 matrices in the cluster for PCC and DCC.

\begin{figure}
\centering
    \includegraphics[width=16.75cm,height=5.75cm]{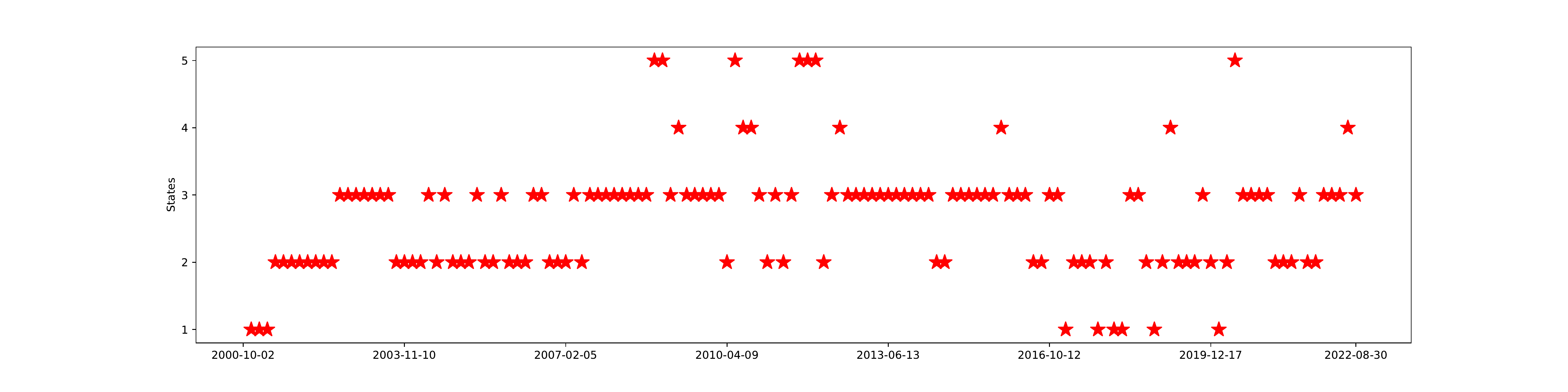} 
    \includegraphics[width=16.75cm,height=5.75cm]{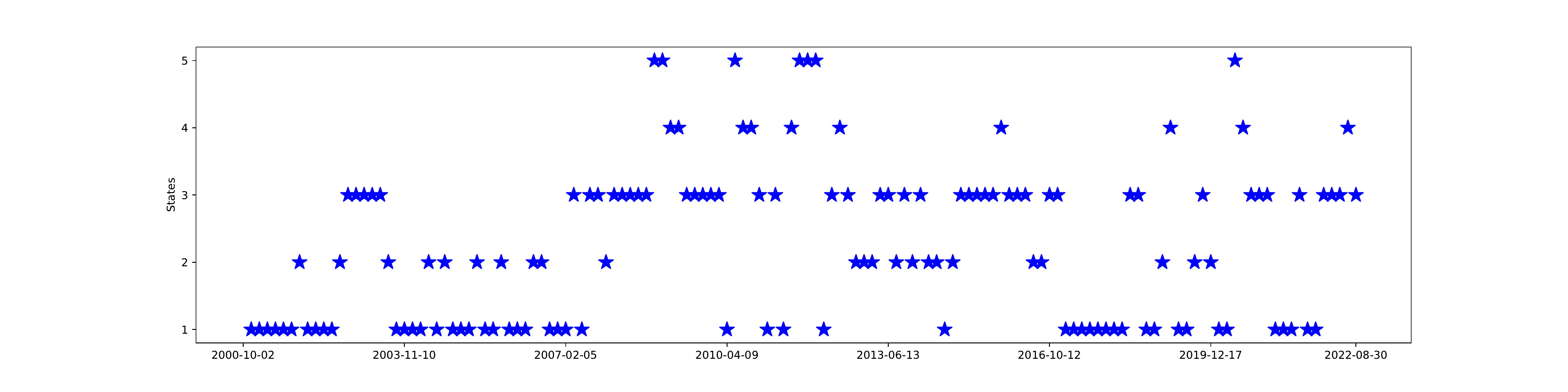} 
    \caption{Dynamical evolution of financial market in time: PCC (top panel) and DCC (bottom panel). The market states $1,2,\ldots,5$ obtained using agglomerative clustering are arranged in increasing order of average correlation coefficients for both PCC and DCC. }
    \label{fig-9}
\end{figure}

\begin{figure}
\centering
    \includegraphics[width=6.75cm,height=6.75cm]{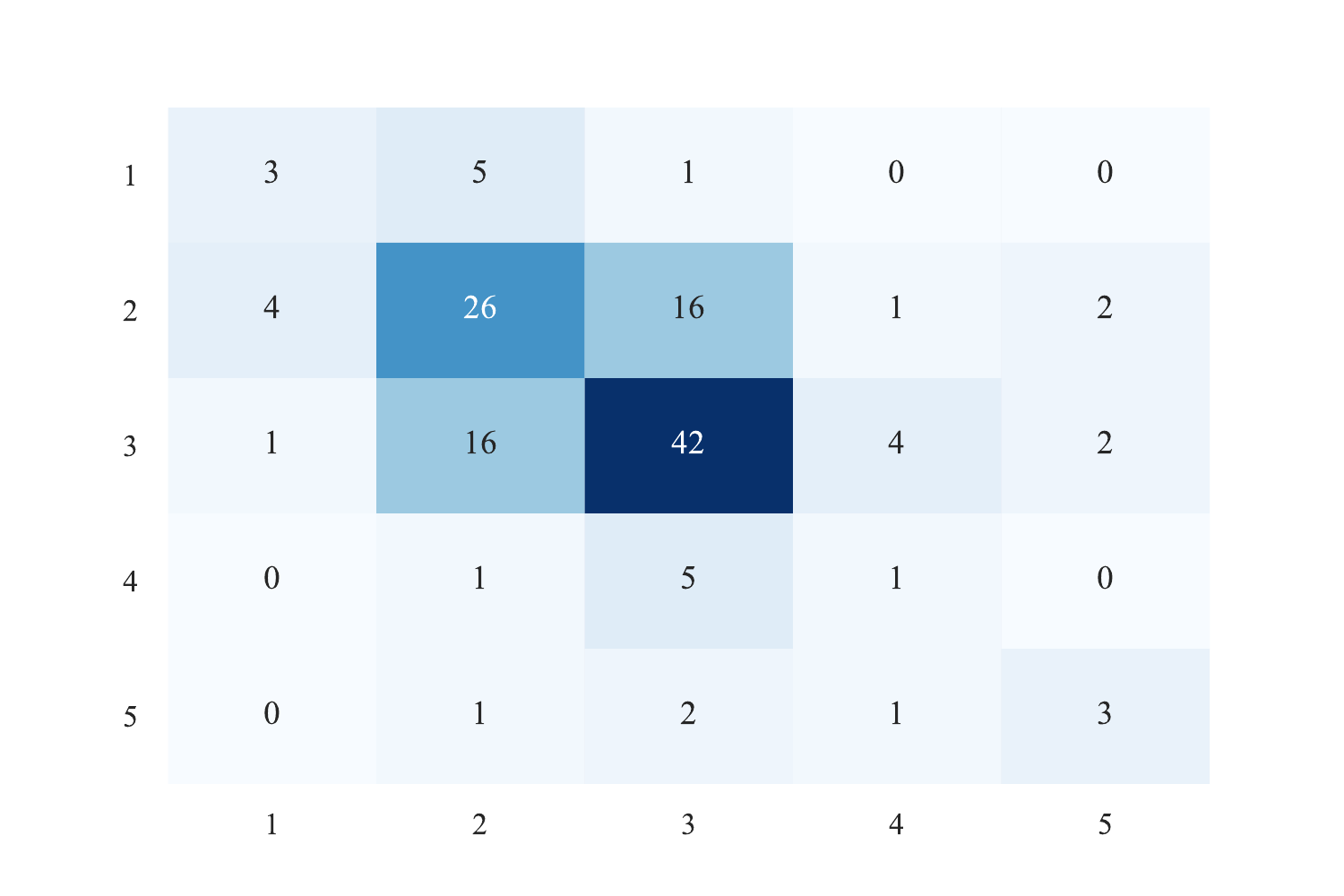} 
    \includegraphics[width=6.75cm,height=6.75cm]{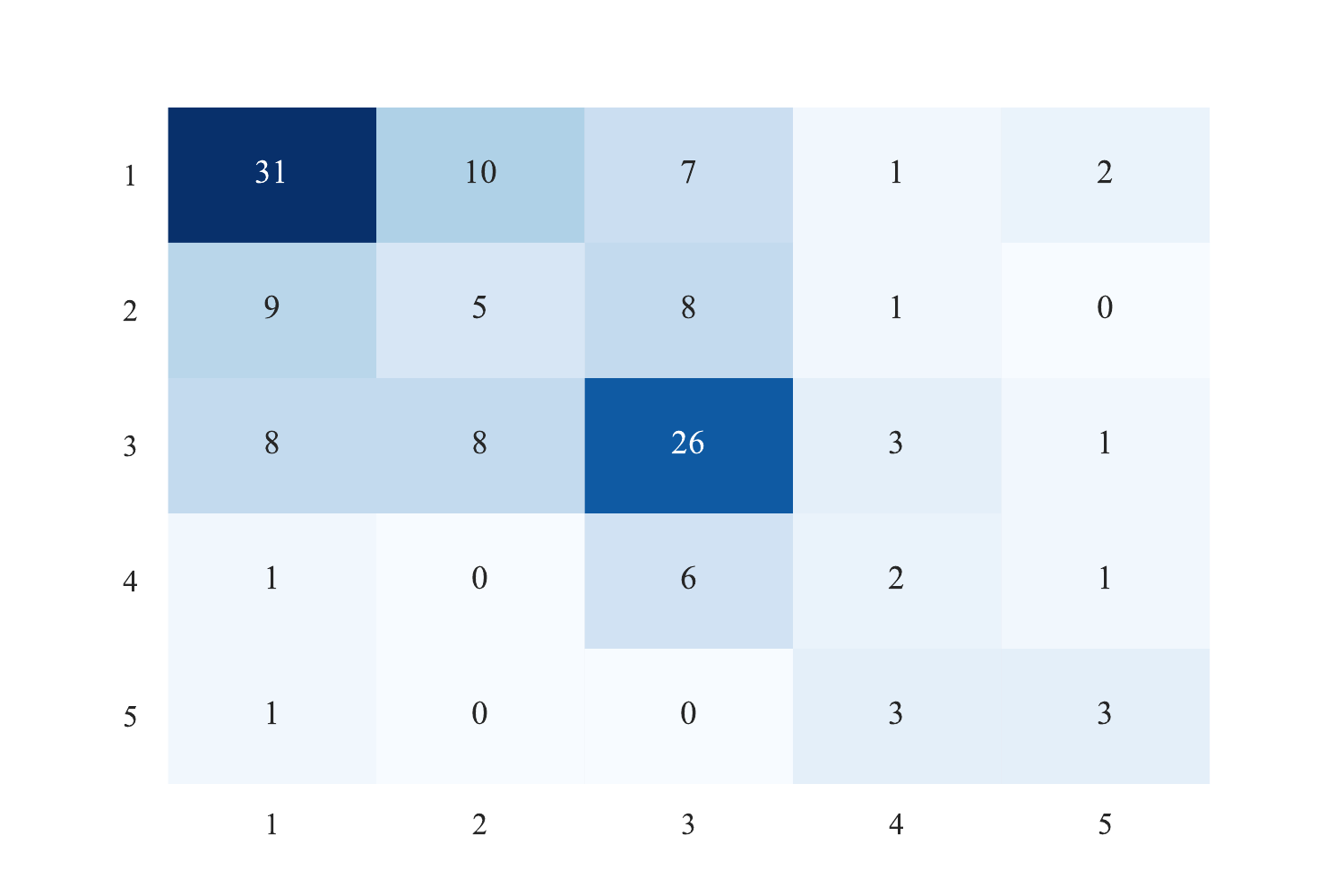} 
    \caption{Transition matrices corresponding to PCC (left panel) and DCC (right panel) showing transition between the five market states obtained using agglomerative clustering.}
    \label{fig-10}
\end{figure}

Dynamical evolution of the financial market can be studied by the transitions between these market states.  The financial market can remain in a particular market state,  can jump to another market state and bounce back or evolve to another market state.  
In Fig.  \ref{fig-9}  the results of the temporal evolution of the market are shown based on both PCC and DCC and Fig. \ref{fig-10} shows the corresponding transition matrices.  
For each market state,  the average correlation coefficients are ordered in ascending order. Transitions are counted when changing epoch, either from one market state to another or if it remained in the same market state.  Most of the values stay close to the diagonal,  this means that the transitions occur in small jumps towards the closest market states or continue in itself and transitions between states with low average correlation and high average correlations are avoided \cite{NJP-18, Gu-22}.

In case of PCC, the state with lowest average correlation (1) never connects to state with highest (5) or second highest (4) average correlation coefficient.  There is a transition from state state 2 to 5 and state 2 to 5 which are indirect transitions as they are in the sequence 1 $\to$ 2 $\to$ 5 and 2 $\to$ 3 $\to$ 5 and these correspond to the crash periods of 2020 and 2011 respectively.  Similarly,  for DCC,  the state with second lowest average correlation (2) never connects to state with highest (5) average correlation coefficient.  However, there are two transitions between 5 and 1 and one transition from 1 to 5.  These correspond to the crash periods of 2010, 2020 and 2011 respectively.  There is also a transition between 1 and 4 that corresponds to the crash period of 2011. This is an indirect one as first transition happens between 1 and 3 and then to 4.  

\section{Conclusions}

We analyzed correlations in S\&P 500 market data for the time period August 2000 to August 2022 using both PCC and DCC.  Notably,  DCC $\geq 0.2$ for the time horizon considered implying that there are non-monotonic correlations present in financial markets at all times. Eigenvalue distributions for DCC and PCC both show a clear shift towards higher values of outliers during the crisis periods of interest (2002, 2008, 2010, 2011, 2015, 2020 and 2022).  The distribution of PR in case of PCC shows a slight upward shift during crisis years of 2008, 2010 and 2011 while we see a slight downward shift in case of DCC during the bubble period of 2002 and crisis years 2008, 2010, 2011 and 2020.  The lesser the average correlation,  prominent is the downward shift in the distribution of PR in case of DCC. 

PCC distinguishes the bubble periods of 2002 and 2007, and the ongoing Russo-Ukrainian war from the crisis periods of 2008, 2010, 2011, 2015 and 2020 depending on kurtosis of the distribution of correlation matrix elements.  DCC distinguishes the bubble periods from the crisis periods depending on skewness of the distribution of correlation matrix elements.  Also,  DCC distinguishes the bubble periods of 2002 and 2007,  the crisis periods of 2008 and 2015, and the ongoing Russo-Ukrainian war from the crisis periods of 2010, 2011 and 2020 depending on kurtosis of the distribution of correlation matrix elements.

Going further, we compare the clustering results for correlation matrices obtained for the selected stocks using PCC and DCC.  We employ agglomerative clustering that uses Euclidean distances and minimizes the sum of squared differences within all clusters.  We obtain five market states corresponding to both PCC and DCC. The crisis periods are in market states with largest and second largest average correlation coefficients. Bubble periods are in the market state with third largest average correlation coefficient.  The two epochs for PCC $\approx 0$ are in the market state with smallest average correlation coefficient; note that this market state has respectively 9 and 51 matrices in the cluster for PCC and DCC.  We also compare the transitions between these market states for both PCC and DCC.  In summary,  results for clustering depend upon the linear (PCC) and non-linear (DCC) nature of the correlation coefficient employed. Preliminary results on financial markets can be viewed in a bachelor thesis \cite{Edu-22}.
 
\acknowledgements

Authors thank Harinder Pal for many useful discussions on clustering algorithms and  help with many figures. Authors acknowledge financial support from CONACYT project Fronteras 10872.

\newpage

\section*{Appendix A: Dendrograms obtained using PCC and DCC}
\label{app-1}

\begin{thebibliography}{99}

\bibitem{Ma-99} R. N. Mantegna and H. E. Stanley,  Introduction to Econophysics: Correlations and Complexity in Finance, (Cambridge University Press,  1999).

\bibitem{Sz-09} G. J.  Sz\'ekely and M. L.  Rizzo, The Annals of Applied Statistics, {\bf 3}, 1236 (2009).

\bibitem{Ede-21} D. Edelmann, T. F.  M\'ori, G. J.  Sz\'ekely,  Statistics and Probability Letters {\bf 169},  108960 (2021).

\bibitem{Brad-81} R. C.  Bradley,  J.  Multivariate Anal.  {\bf 11}, 1 (1981).

\bibitem{Brad-88}  R. C.  Bradley,  Ann.  Probab.  {\bf 16},  313 (1988).

\bibitem{Brad-07}  R. C.  Bradley,  Introduction to Strong Mixing Condition,  {\bf 1–3},  (Kendrick Press, , Heber City (Utah), 2007). 

\bibitem{Sz-08} G. J.  Sz\'ekely and N. K. Bakirov,  Brownian covariance and CLT for stationary sequences, Technical Report No. 08-01,  Dept. Mathematics and Statistics,  Bowling Green State Univ.,  Bowling Green, OH (2008).

\bibitem{Gu-98} T.  Guhr,  A. Mueller,  H. A.  Weidenmueller,  Physics Reports {\bf 299}, 189 (1998).

\bibitem{Pl-99} V.  Plerou, P.  Gopikrishnan,  B. Rosenow,  L.  A.  Nunes Amaral,  and H. E. Stanley,  Phys.  Rev.  Lett.  {\bf 83}, 1471 (1999).

\bibitem{Mu-12} M. C.  M\"unnix {\it et.  al. },  Scientific Reports {\bf 2},  644 (2012).

\bibitem{Gu-15} D. Chetalova, R Sch\"afer, and T. Guhr,  J.  Stat.  Mech.  {\bf 2015},  P01029 (2015).

\bibitem{NJP-18} H. K.  Pharasi {\it et.  al. },  New Journal of Physics {\bf 20}, 103041 (2018).

\bibitem{Plos-15} N. Musmeci, T.  Aste,  and T. Di Matteo,  PLoS ONE {\bf 10}, 1 (2015).

\bibitem{YF} Yahoo finance database, https://finance.yahoo.com/, accessed on 10 October,  2022 for S\&P 500.

\bibitem{Ed-88} A.  Edelman,  SIAM Journal on Matrix Analysis and Applications, {\bf 9}, 543 (1988).

\bibitem{La-99} L.  Laloux,  P.  Cizeau,  J. -P.  Bouchaud,  and M. Potters,  Phys.  Rev.  Lett.  {\bf 83}, 1467 (1999).

\bibitem{Ma-18} M.  Vyas, T.  Guhr, and T.  H.  Seligman, Scientific reports {\bf 8}, 1 (2018).

\bibitem{Deo-19} P.  Bhadola and N.  Deo,  "Spectral and Network Method in Financial Time Series Analysis: A Study on Stock and Currency Market",  in A. S. Chakrabarti et al. (eds.),  Network Theory and Agent-Based Modeling in Economics and Finance (2019) pp. 331-352.

\bibitem{Pha-RMT} H.  K.  Pharasi,  K.  Sharma,  A.  Chakraborti, and T.  H.  Seligman, "Complex market dynamics in the light of random matrix theory", in New Perspectives and Challenges in Econophysics and Sociophysics, edited by F. Abergel, B. K. Chakrabarti, A. Chakraborti, N. Deo, and K. Sharma (Springer International Publishing, Cham, 2019) pp. 13–34.

\bibitem{Ko-book} V. K. B.  Kota,  Embedded Random Matrix Ensembles in Quantum Physics (Springer, Heidelberg, 2014).

\bibitem{St-87} A.  Stuart and J.  K. Ord,  Kendall’s Advanced Theory of Statistics : Distribution Theory (Oxford University Press, New York, 1987).

\bibitem{Gu-22} A. J. Heckens and T. Guhr,  J. Stat. Mech.  {\bf 2022},  043401 (2022) 

\bibitem{Edu-22} J.  E.  Salgado-Hern\'andez,  (Licenciatura thesis, UNAM) {\it Correlaci\'on y agrupaciones de series de tiempo financieras} (2023).

\end{thebibliography}
\end{document}